\documentclass[10pt,a4papper]{article}
\usepackage[utf8]{inputenc}
\usepackage{indentfirst}
\usepackage[]{graphicx}
\usepackage{geometry} 
\begin{document}
	\title{\textbf{The decay $\tau \to K^{-} \eta \nu_{\tau}$ in the extended Nambu-Jona-Lasinio model taking into account the meson interaction in the final state}}
	\author{M. K. Volkov\footnote{volkov@theor.jinr.ru}, A. A. Pivovarov\footnote{tex$\_$k@mail.ru}\\
		\small
		\emph{BLTP, Joint Institute for Nuclear Research, Dubna, 141980, Russia}}
	\date{}
	\maketitle
	\small
	
\begin{abstract}
	The branching fraction of the decay $\tau \to K^{-} \eta \nu_{\tau}$ has been calculated by using the extended Nambu-Jona-Lasinio model taking into account the meson interaction in the final state. The contact term and the channel with the intermediate ground $K^{*}(892)$ and first radially excited state $K^{*}(1410)$ have been calculated. For the purpose of describing the meson interaction in the final state, the triangle diagram with three mesons has been considered. The cutoff parameter of divergent integrals of the meson loop is equal to the appropriate parameter used in our previous work to describe the decay $\tau \to K^{-} \pi^{0} \nu_{\tau}$. Thereby, the present work does not set any additional arbitrary parameters. The obtained results are in satisfactory agreement with the experimental data.
\end{abstract}

\large
\section{Introduction}
Numerous calculations have shown that the Nambu-Jona-Lasinio (NJL) model \cite{Volkov:1986zb, Volkov:1993jw, Volkov:2005kw, Volkov:2017arr} allows a satisfactory description of the main tau lepton decays. In the framework of this model, a series of tau lepton decays was described in satisfactory agreement with the experimental data: $\tau \to M \nu_{\tau}$, $\tau \to V P \nu_{\tau}$, where $V$ is a vector meson, $P$ is a pseudoscalar meson, $M$ is a vector or pseudoscalar meson \cite{Volkov:2017arr, Volkov:2017cmv, Volkov:2019yhy, Volkov:2019cja, Volkov:2019udu, Volkov:2019jug, Volkov:2019yli, Volkov:2019izp}. However, several tau decays with pseudoscalar mesons in the final state can not be described in the framework of the NJL model in agreement with the experimental data. This may be caused by the necessity of taking into account the mesons interaction in the final state, which requires going beyond the $1/N_{c}$ approximation order that the NJL model has been formulated in. The final state interaction with violation of this approximation was considered in recent works describing the processes $\tau^{-} \to \pi^{-} \pi^{0} \nu_{\tau}$ \cite{Volkov:2020dvz} and $\tau^{-} \to K^{-} \pi^{0} \nu_{\tau}$ \cite{Kpi}. The energy threshold of the final meson production in these processes is lower than the mass values of the intermediate ground states $\rho(770)$ and $K^{*}(892)$, respectively. Therefore, the excited states do not make a significant contribution. In contrast, the energy threshold of the process $\tau^{-} \to K^{-} \eta \nu_{\tau}$ is higher than the mass value of the ground state $K^{*}(892)$. As a result, this process is sensitive to the resonance $K^{*}(1410)$ that plays a more significant role. Thus, this process having a similar structure to the process $\tau^{-} \to K^{-} \pi^{0} \nu_{\tau}$ is more useful for studying the excited meson states.

In the present work, we consider the process $\tau^{-} \to K^{-} \eta \nu_{\tau}$ taking into account the meson interaction in the final state. It is similar to the decay $\tau^{-} \to K^{-} \pi^{0} \nu_{\tau}$. However, due to a significant contribution of the resonance $K^{*}(1410)$, the extended NJL model \cite{Volkov:1996br, Volkov:1996fk} is applied to calculate it. This model allows one to describe the first radially excited meson states without violation of the $U(3) \times U(3)$ chiral symmetry. To take into account the meson interaction in the final state, we are going beyond this model as well. To describe this interaction one triangle diagram with the exchange of the charged vector meson $K^{*\pm}(892)$ is required. However, the meson vertices in this triangle are more complicated due to the structure of the $\eta$ meson. It is interesting that the cutoff values of the meson triangles of the processes $\tau^{-} \to K^{-} \eta \nu_{\tau}$ and $\tau^{-} \to K^{-} \pi^{0} \nu_{\tau}$ are equal.

Experimentally, the process $\tau^{-} \to K^{-} \eta \nu_{\tau}$ was initially studied by the CLEO~\cite{Bartelt:1996iv} and ALEPH~\cite{Buskulic:1996qs} collaborations. Later, the Belle~\cite{Inami:2008ar} and Babar~\cite{delAmoSanchez:2010pc} collaborations performed more precise measurements of this decay.

This process was also studied theoretically. For instance, in the work \cite{Li:1996md}, it was considered by using the vector meson dominance model and the effective chiral theory. In \cite{Escribano:2013bca}, it was treated within the chiral perturbation theory with resonances.

\section{The Lagrangian of the extended NJL model}
A fragment of the chiral quark-meson Lagrangian of the extended NJL model containing the necessary vertices takes the form~\cite{Volkov:2017arr}:
\begin{eqnarray}
		\Delta L_{int} & = &
		\bar{q} \left[ \frac{1}{2} \gamma^{\mu} \sum_{j = \pm}\lambda_{j}^{K} \left(A_{K^{*}}K^{*j}_{\mu} + B_{K^{*}}K^{*'j}_{\mu}\right) \right. \nonumber\\
		&& + \left. i \gamma^{5} \sum_{j = \pm} \lambda_{j}^{K} \left(A_{K}K^{j} + B_{K}K'^{j}\right) + i \gamma^{5} \sum_{j = u,s} \lambda_{j}^{\eta} A_{\eta^{j}}\eta^{j}\right]q,
	\end{eqnarray}
where $q$ and $\bar{q}$ are the u, d and s quark fields with the constituent masses $m_{u} \approx m_{d} = 280$~MeV, $m_{s} = 420$~MeV, the excited states are marked with a prime. The factors $A$ and $B$ for the strange mesons take the form:
\begin{eqnarray}
\label{verteces1}
	A_{M} & = & \frac{1}{\sin(2\theta_{M}^{0})}\left[g_{M}\sin(\theta_{M} + \theta_{M}^{0}) +
	g_{M}^{'}f_{M}(k_{\perp}^{2})\sin(\theta_{M} - \theta_{M}^{0})\right], \nonumber\\
	B_{M} & = & \frac{-1}{\sin(2\theta_{M}^{0})}\left[g_{M}\cos(\theta_{M} + \theta_{M}^{0}) +
	g_{M}^{'}f_{M}(k_{\perp}^{2})\cos(\theta_{M} - \theta_{M}^{0})\right].
\end{eqnarray}
The index $M$ designates the appropriate meson.
	
The factor $A$ for the $\eta$ meson takes a different form. This is due to the mixing of four states in the case of the $\eta$ meson in contrast to the mixing of two states in the case of strange mesons \cite{Volkov:2017arr}:
\begin{eqnarray}
\label{verteces2}
	A_{\eta^{u}} & = & 0.71 g_{\eta^{u}} + 0.11 g_{\eta^{u}}^{'} f_{uu}(k_{\perp}^{2}), \nonumber\\
	A_{\eta^{s}} & = & 0.62 g_{\eta^{s}} + 0.06 g_{\eta^{s}}^{'} f_{ss}(k_{\perp}^{2}).
\end{eqnarray}
	
$f\left(k_{\perp}^{2}\right) = \left(1 + d k_{\perp}^{2}\right)\Theta(\Lambda^{2} - k_{\perp}^2)$ is the form factor describing the first radially excited meson states. The slope parameter $d$ has been fixed by the requirement of quark condensate invariability after including the radially excited states. It depends only on the quark composition of the appropriate meson:
\begin{eqnarray}
	&d_{uu} = -1.784 \times 10^{-6} \textrm{MeV}^{-2}, \quad d_{us} = -1.761 \times 10^{-6}\textrm{MeV}^{-2}.& \nonumber\\
	&d_{ss} = -1.737 \times 10^{-6} \textrm{MeV}^{-2}.&
\end{eqnarray}
	
The transverse relative momentum of the inner quark-antiquark system can be represented as follows:
\begin{eqnarray}
	k_{\perp} = k - \frac{(kp) p}{p^2},
\end{eqnarray}
where $p$ is a meson momentum. In the rest frame of a meson
\begin{eqnarray}
	k_{\perp} = (0, {\bf k}).
\end{eqnarray}
Therefore, this momentum can be applied in the three-dimensional form.
	
The parameters $\theta_{M}$ are the mixing angles appearing as a result of the diagonalization of the Lagrangian with the ground and first radially excited meson states \cite{Volkov:2017arr, Volkov:1996fk}:
\begin{eqnarray}
\label{angels}
	&\theta_{K} = 58.11^{\circ}, \quad \theta_{K^{*}} = 84.74^{\circ}.&
\end{eqnarray}
The auxiliary values $\theta_{M}^{0}$ have been included for convenience:
\begin{eqnarray}
\label{tetta0}
	&\sin\left(\theta_{M}^{0}\right) = \sqrt{\frac{1 + R_{M}}{2}},& \nonumber\\
	&R_{K^{*}} = \frac{I_{11}^{f_{us}}}{\sqrt{I_{11}I_{11}^{f^{2}_{us}}}}, \quad R_{K} = \frac{I_{11}^{f_{us}}}{\sqrt{Z_{K}I_{11}I_{11}^{f^{2}_{us}}}}, & 
\end{eqnarray}
where
\begin{eqnarray}
\label{Zk}
	Z_{K} & = & \left(1 - \frac{3}{2} \frac{(m_{u} + m_{s})^{2}}{M_{K_{1A}}^{2}}\right)^{-1}, \nonumber\\
	M_{K_{1A}} & = & \left(\frac{\sin^{2}{\alpha}}{M^{2}_{K_{1}(1270)}} + \frac{\cos^{2}{\alpha}}{M^{2}_{K_{1}(1400)}}\right)^{-1/2}.
\end{eqnarray}

Here $Z_{K}$ is the additional renormalization constant appearing in the $K-K_{1}$ transitions, $M_{K_{1}(1270)} = 1253 \pm 7$~MeV is the mass of the meson $K_{1}(1270)$, $M_{K_{1}(1400)} = 1403 \pm 7$~MeV is the mass of the meson $K_{1}(1400)$ \cite{Zyla:2020zbs}. In the expression for $Z_{K}$, the splitting of the state $K_{1A}$ into two physical mesons $K_{1}(1270)$ and $K_{1}(1400)$ with the mixing angle $\alpha = 57^{\circ}$ \cite{Volkov:2019yhy} is taken into account.
	
The integrals appearing in the quark loops as a result of the renormalization of the Lagrangian:
\begin{eqnarray}
	I_{n_{1}n_{2}}^{f^{m}} =
	-i\frac{N_{c}}{(2\pi)^{4}}\int\frac{f^{m}({\bf k}^{2})}{(m_{u}^{2} - k^2)^{n_{1}}(m_{s}^{2} - k^2)^{n_{2}}}\Theta(\Lambda^{2} - {\bf k}^2)
	\mathrm{d}^{4}k,
\end{eqnarray}
where $\Lambda = 1.03$~GeV is the three-dimensional cutoff parameter.
	
Then
\begin{eqnarray}
	&\theta_{K}^{0} = 55.52^{\circ}, \quad \theta_{K^{*}}^{0} = 59.56^{\circ}.& 
\end{eqnarray}
	
The matrices $\lambda$ are the linear combinations of the Gell-Mann matrices:
\begin{eqnarray}
	\lambda_{+}^{K} = \frac{\lambda_{4} + i\lambda_{5}}{\sqrt{2}}, &\quad&
	\lambda_{-}^{K} = \frac{\lambda_{4} - i\lambda_{5}}{\sqrt{2}}, \nonumber\\
	\lambda_{u}^{\eta} = \frac{2\lambda_{0} + \sqrt{3}\lambda_{8}}{3}, &\quad&
	\lambda_{s}^{\eta} = \frac{-\sqrt{2}\lambda_{0} + \sqrt{6}\lambda_{8}}{3},
\end{eqnarray}
where $\lambda_{0}$ is the unit matrix.
	
The coupling constants:
\begin{eqnarray}
	\label{Couplings}
	g_{K^{*}} = \left(\frac{2}{3}I_{11}\right)^{-1/2}, &\quad& g_{K^{*}}^{'} = \left(\frac{2}{3}I_{11}^{f^{2}}\right)^{-1/2},  \nonumber\\
	g_{K} =  \left(\frac{4}{Z_{K}}I_{11}\right)^{-1/2}, &\quad& g_{K}^{'} =  \left(4I_{11}^{f^{2}}\right)^{-1/2}, \nonumber\\
	g_{\eta^{u}} =  \left(\frac{4}{Z_{\eta^{u}}}I_{20}\right)^{-1/2}, &\quad& g_{\eta^{u}}^{'} =  \left(4 I_{20}^{f^{2}}\right)^{-1/2}, \nonumber\\
	g_{\eta^{s}} =  \left(\frac{4}{Z_{\eta^{s}}}I_{02}\right)^{-1/2}, &\quad& g_{\eta^{s}}^{'} =  \left(4 I_{02}^{f^{2}}\right)^{-1/2},
\end{eqnarray}
where
\begin{eqnarray}
	Z_{\eta^{u}} & \approx & Z_{\pi} = \left(1 - 6\frac{m_{u}^{2}}{M^{2}_{a_{1}}}\right)^{-1}, \nonumber\\
	Z_{\eta^{s}} & = & \left(1 - 6\frac{m_{s}^{2}}{M^{2}_{f_{1}^{s}}}\right)^{-1}.
\end{eqnarray}

Here $Z_{\eta^{s}}$ is the additional renormalization constant appearing in the transitions between the axial vector meson $f_{1}(1420)$ and the $s$ quark part of the $\eta$ meson, $M_{f_{1}^{s}} = 1426$~MeV \cite{Zyla:2020zbs}. The parameter $Z_{\eta^{u}}$ in the NJL model is usually taken approximately equal to the parameter $Z_{\pi}$ appearing in the $a_{1}-\pi$ transitions; $M_{a_{1}} = 1230 \pm 40$~MeV is the mass of the meson $a_{1}(1260)$ \cite{Zyla:2020zbs}.

\section{The process $\tau^{-} \to K^{-} \eta \nu_{\tau}$ in the extended NJL model}
\begin{figure}[h]
	\center{\includegraphics[scale = 0.8]{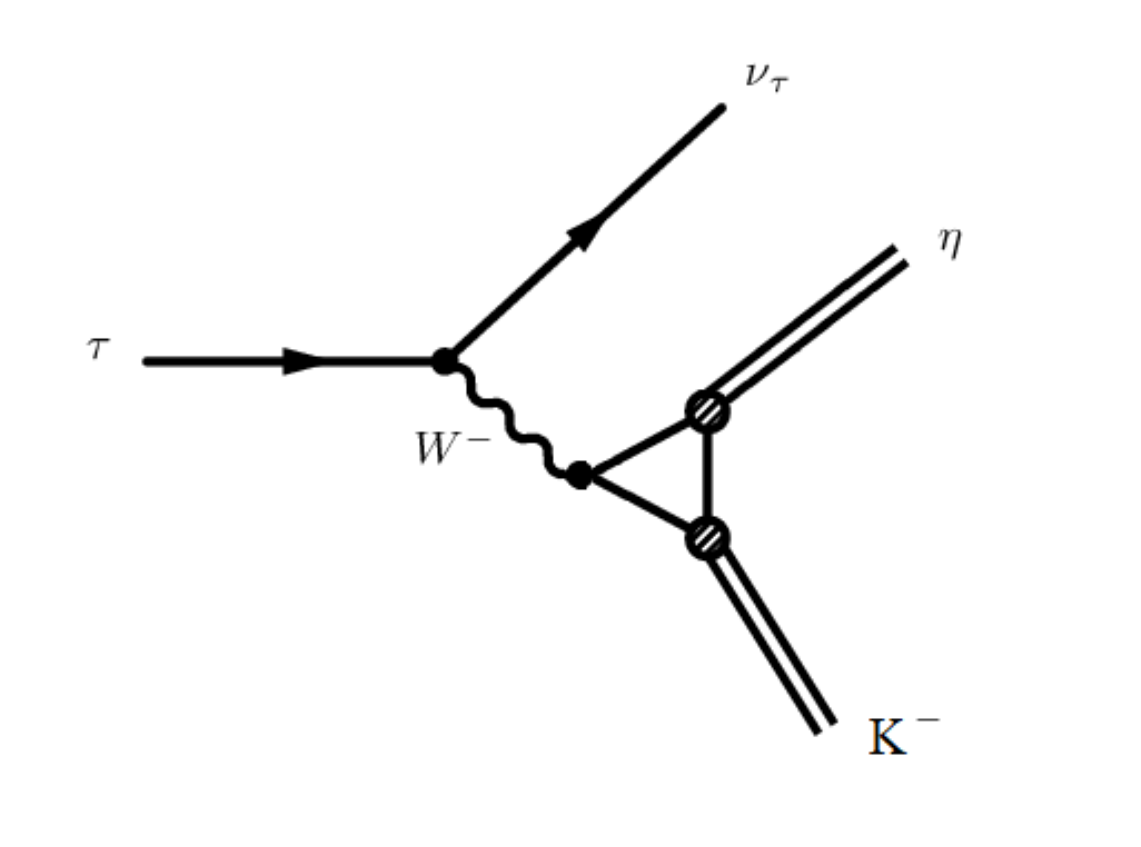}}
	\caption{The contact diagram.}
	\label{Contact}
\end{figure}

\begin{figure}[h]
	\center{\includegraphics[scale = 1]{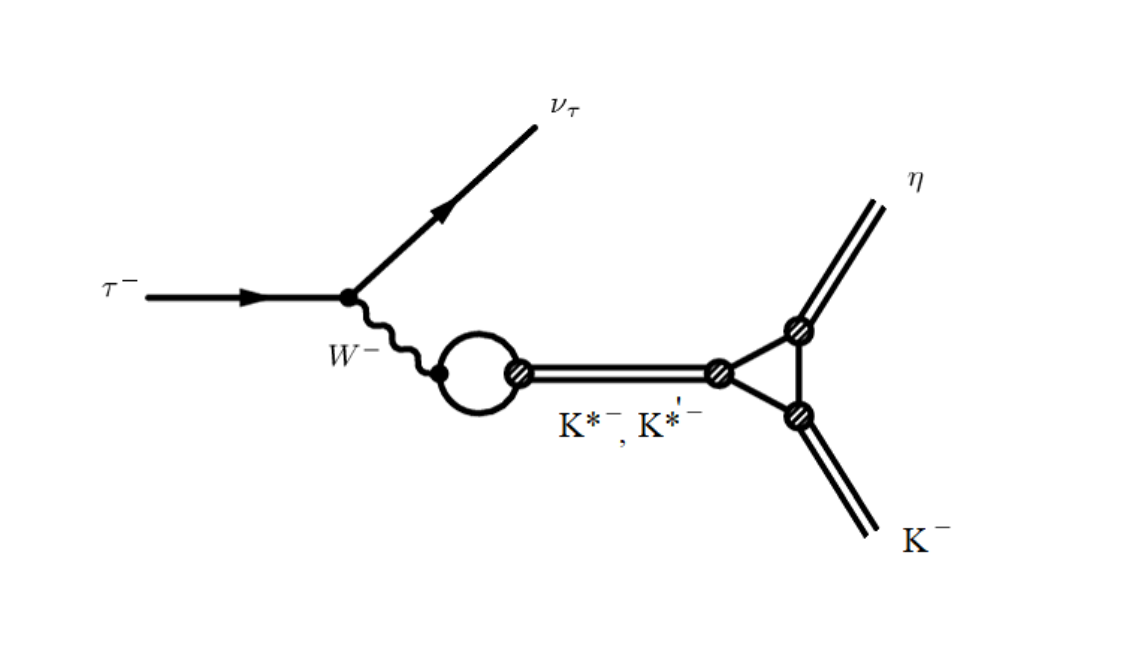}}
	\caption{The diagram with the intermediate mesons.}
	\label{Interm}
\end{figure}

The process $\tau^{-} \to K^{-} \eta \nu_{\tau}$ in the extended NJL model can be described by the diagrams shown in Figs.~\ref{Contact},~\ref{Interm}.

The amplitude of this process takes the form:
\begin{eqnarray}
    M_{tree} & = &-2 G_{f} V_{us} \left(I_{11}^{K \eta^{u}} + \sqrt{2} I_{11}^{K \eta^{s}}\right) L_{\mu} \left[\left(T_{K}^{(c)}p_{K} - T_{\eta}^{(c)}p_{\eta}\right)^{\mu} \right. \nonumber\\
    && + \frac{C_{K^{*}}}{g_{K^{*}}} \frac{I_{11}^{K K^{*} \eta^{u}} + \sqrt{2} I_{11}^{K K^{*} \eta^{s}}}{I_{11}^{K \eta^{u}} + \sqrt{2} I_{11}^{K \eta^{s}}} \frac{g^{\mu\nu} q^{2} f(q^{2}) - q^{\mu}q^{\nu}f(M_{K^{*}}^{2})}{M_{K^{*}}^{2} - q^{2} - i \sqrt{q^{2}}\Gamma_{K^{*}}} \left(T_{K}^{(K^{*})}p_{K} - T_{\eta}^{(K^{*})}p_{\eta}\right)_{\nu} \nonumber\\
    && \left. + \frac{C_{K^{*'}}}{g_{K^{*}}} \frac{I_{11}^{K K^{*'} \eta^{u}} + \sqrt{2} I_{11}^{K K^{*'} \eta^{s}}}{I_{11}^{K \eta^{u}} + \sqrt{2} I_{11}^{K \eta^{s}}} \frac{g^{\mu\nu} q^{2} f(q^{2}) - q^{\mu}q^{\nu}f(M_{K^{*'}}^{2})}{M_{K^{*'}}^{2} - q^{2} - i \sqrt{q^{2}}\Gamma_{K^{*'}}} \left(T_{K}^{(K^{*'})}p_{K} - T_{\eta}^{(K^{*'})}p_{\eta}\right)_{\nu} \right],
\end{eqnarray}
where $G_{f}$ is the Fermi constant, $V_{us}$ is the element of the Cabibbo–Kobayashi–Maskawa matrix, $L_{\mu}$ is the lepton current, the expression for $f(q^{2})$ takes the form:
\begin{eqnarray}
    f(q^{2}) = 1 - \frac{3}{2} \frac{\left(m_{s} - m_{u}\right)^{2}}{q^{2}}.
\end{eqnarray}

The factors $T$ describe the transitions between the axial vector and pseudoscalar mesons:
\begin{eqnarray}
    T_{K}^{(c)} & = & 1 - 2 \frac{m_{s} I_{11}^{K_{1} \eta^{u}} + \sqrt{2} m_{u} I_{11}^{K_{1} \eta^{s}}}{I_{11}^{K \eta^{u}} + \sqrt{2} I_{11}^{K \eta^{s}}} I_{11}^{K_{1} K} \frac{m_{s} + m_{u}}{M_{K_{1A}}^{2}}, \nonumber\\
    T_{\eta}^{(c)} & = & 1 - 2 \frac{I_{11}^{K f^{u}} I_{20}^{f^{u} \eta^{u}}}{I_{11}^{K \eta^{u}} + \sqrt{2} I_{11}^{K \eta^{s}}} \frac{m_{u}\left(3m_{u} - m_{s}\right)}{M_{f_{1}^{u}}^{2}} - 2 \sqrt{2} \frac{I_{11}^{K f^{s}} I_{02}^{f^{s} \eta^{s}}}{I_{11}^{K \eta^{u}} + \sqrt{2} I_{11}^{K \eta^{s}}} \frac{m_{s}\left(3m_{s} - m_{u}\right)}{M_{f_{1}^{s}}^{2}}, \nonumber\\
    T_{K}^{(K^{*})} & = & 1 - 2 \frac{m_{s} I_{11}^{K^{*} K_{1} \eta^{u}} + \sqrt{2} m_{u} I_{11}^{K^{*} K_{1} \eta^{s}}}{I_{11}^{K K^{*} \eta^{u}} + \sqrt{2} I_{11}^{K K^{*} \eta^{s}}} I_{11}^{K_{1} K} \frac{m_{s} + m_{u}}{M_{K_{1A}}^{2}}, \nonumber\\
    T_{\eta}^{(K^{*})} & = & 1 - 2 \frac{I_{11}^{K^{*} K f^{u}} I_{20}^{f^{u} \eta^{u}}}{I_{11}^{K^{*} K \eta^{u}} + \sqrt{2} I_{11}^{K^{*} K \eta^{s}}} \frac{m_{u}\left(3m_{u} - m_{s}\right)}{M_{f_{1}^{u}}^{2}} - 2 \sqrt{2} \frac{I_{11}^{K^{*} K f^{s}} I_{02}^{f^{s} \eta^{s}}}{I_{11}^{K^{*} K \eta^{u}} + \sqrt{2} I_{11}^{K^{*} K \eta^{s}}} \frac{m_{s}\left(3m_{s} - m_{u}\right)}{M_{f_{1}^{s}}^{2}}, \nonumber\\
    T_{K}^{(K^{*'})} & = & 1 - 2 \frac{m_{s} I_{11}^{K^{*'} K_{1} \eta^{u}} + \sqrt{2} m_{u} I_{11}^{K^{*'} K_{1} \eta^{s}}}{I_{11}^{K K^{*'} \eta^{u}} + \sqrt{2} I_{11}^{K K^{*'} \eta^{s}}} I_{11}^{K_{1} K} \frac{m_{s} + m_{u}}{M_{K_{1A}}^{2}}, \nonumber\\
    T_{\eta}^{(K^{*'})} & = & 1 - 2 \frac{I_{11}^{K^{*'} K f^{u}} I_{20}^{f^{u} \eta^{u}}}{I_{11}^{K^{*'} K \eta^{u}} + \sqrt{2} I_{11}^{K^{*'} K \eta^{s}}} \frac{m_{u}\left(3m_{u} - m_{s}\right)}{M_{f_{1}^{u}}^{2}} - 2 \sqrt{2} \frac{I_{11}^{K^{*'} K f^{s}} I_{02}^{f^{s} \eta^{s}}}{I_{11}^{K^{*'} K \eta^{u}} + \sqrt{2} I_{11}^{K^{*'} K \eta^{s}}} \frac{m_{s}\left(3m_{s} - m_{u}\right)}{M_{f_{1}^{s}}^{2}}.
\end{eqnarray}

The integrals with the vertices from the Lagrangian in the numerators applied in the amplitude are:
\begin{eqnarray}
	I_{n_{1} n_{2}}^{M, \dots, M^{'}, \dots} =
	-i\frac{N_{c}}{(2\pi)^{4}}\int\frac{A_{M} \dots B_{M} \dots}{(m_{u}^{2} - k^2)^{n_{1}}(m_{s}^{2} - k^2)^{n_{2}}} \theta(\Lambda^{2} - {\bf k}^2)
	\mathrm{d}^{4}k,
\end{eqnarray}
where $A_{M}, B_{M}$ are defined in (\ref{verteces1}) and (\ref{verteces2}).

The constants $C_{K^{*}}$ and $C_{K^{*'}}$ 
\begin{eqnarray}
    C_{K^{*}} & = & \frac{1}{\sin\left(2\theta_{K^{*}}^{0}\right)} \left[\sin\left(\theta_{K^{*}} + \theta_{K^{*}}^{0}\right) + R_{K^{*}}\sin\left(\theta_{K^{*}} - \theta_{K^{*}}^{0}\right)\right], \nonumber\\
    C_{K^{*'}} & = & \frac{-1}{\sin\left(2\theta_{K^{*}}^{0}\right)} \left[\cos\left(\theta_{K^{*}} + \theta_{K^{*}}^{0}\right) + R_{K^{*}}\cos\left(\theta_{K^{*}} - \theta_{K^{*}}^{0}\right)\right]
\end{eqnarray}
appear in the transitions between the W boson and the intermediate vector meson. Here $\theta$ is the mixing angle of the ground and first radially excited meson states defined in (\ref{angels}). The values $\theta^{0}$ and $R$ are defined in (\ref{tetta0}).
	
The first term of the amplitude in the squared brackets describes the contact diagram. The second and the third terms describe the diagrams with the intermediate ground and the excited vector mesons $K^{*}(892)$ and $K^{*}(1410)$.

This amplitude leads to the following branching fraction of this decay:
\begin{eqnarray}
	Br(\tau^{-} \to K^{-} \eta \nu_{\tau})_{tree} & = & 1.35  \times 10^{-4}.
\end{eqnarray}

The experimental value \cite{Zyla:2020zbs} is:
\begin{eqnarray}
	Br(\tau^{-} \to K^{-} \eta \nu_{\tau})_{exp} & = & (1.55 \pm 0.08) \times 10^{-4}.
\end{eqnarray}

As one can see, the obtained result is lower than the experimental value. This could mean that additional effects should be taken into account, such as the meson interaction in the final state.

\section{The process $\tau^{-} \to K^{-} \eta \nu_{\tau}$ taking into account the meson interaction in the final state}
\begin{figure}[h]
	\center{\includegraphics[scale = 0.8]{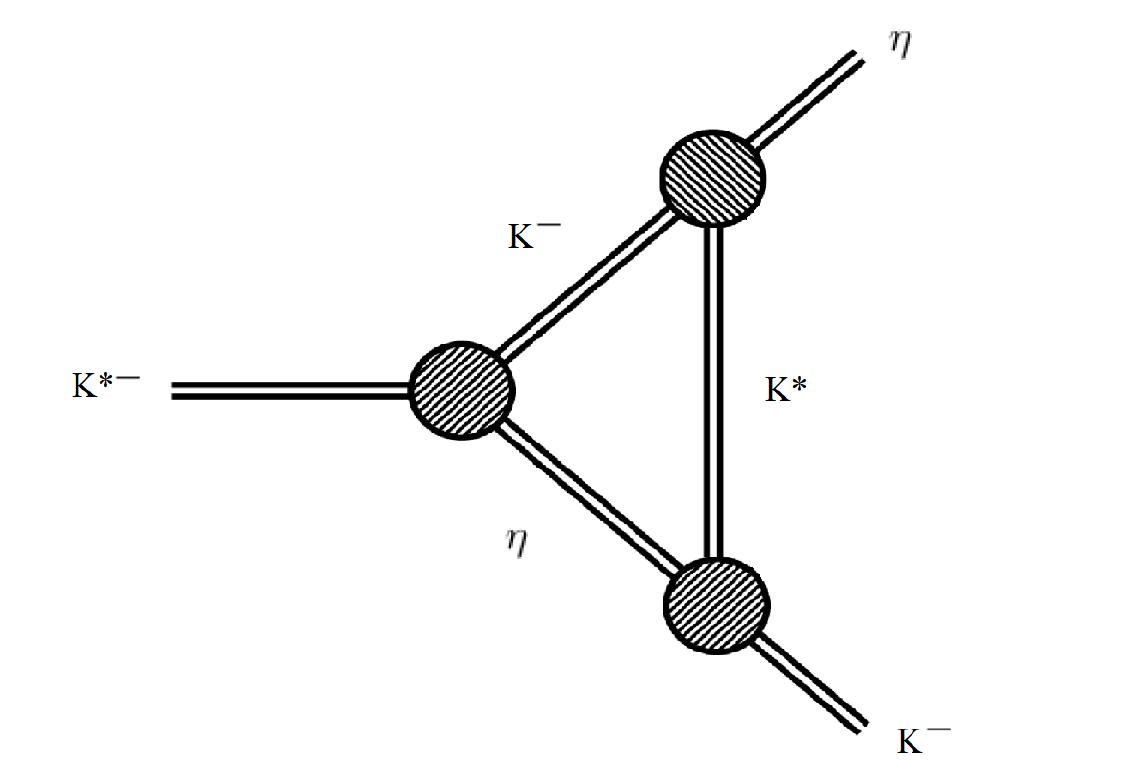}}
	\caption{The meson triangle.}
	\label{Triangle}
\end{figure}

The interaction in the final state in this process can be taken into account through the exchange by the charged meson $K^{*}$ between the final kaon and $\eta$ meson leading to their replacing. As a result, the triangle shown in Fig.~\ref{Triangle} appears.

The meson vertex of the Lagrangian that is necessary for building this triangle in the extended NJL model takes the form:
\begin{eqnarray}
    -i 2 \left(I_{11}^{K K^{*} \eta^{u}} + \sqrt{2} I_{11}^{K K^{*} \eta^{s}}\right) K_{\mu}^{*-} \left(T_{K}^{(K^{*})} \partial^{\mu}K^{+}\eta - T_{\eta}^{(K^{*})}K^{+}\partial^{\mu}\eta\right).
\end{eqnarray}

This meson triangle leads to the integral
\begin{eqnarray}
   &F_{\mu} =  \int \frac{\left(T_{K}^{(K^{*})}k - \left(T_{K}^{(K^{*})} + T_{\eta}^{(K^{*})}\right)p_{\eta}\right)_{\lambda}\left(T_{\eta}^{(K^{*})}k + \left(T_{K}^{(K^{*})} + T_{\eta}^{(K^{*})}\right)p_{K}\right)_{\nu}
	\left(\left(T_{K}^{(K^{*})} + T_{\eta}^{(K^{*})}\right)k + T_{\eta}^{(K^{*})}p_{K} - T_{K}^{(K^{*})}p_{\eta}\right)_{\mu} \left(g^{\nu\lambda} 
	- \frac{k^{\nu}k^{\lambda}}{M_{K^{*}}^{2}}\right)}{\left[k^{2} - M_{K^{*}}^{2}\right]
	\left[(k + p_{K})^{2} - M_{\eta}^{2}\right]\left[(k - p_{\eta})^{2} - M_{K}^{2}\right]}& \nonumber\\
	&\times \frac{d^{4}k}{(2\pi)^{4}}.&
\end{eqnarray}

As one can see, this integral is similar to the one for the process $\tau \to K \pi \nu_{\tau}$ \cite{Kpi}.

The amplitude of the additional contribution of the meson loop for the process $\tau^{-} \to K^{-} \eta \nu_{\tau}$ takes the form:
\begin{eqnarray}
    M_{loop} & = & 8i G_{f} V_{us} \left(I_{11}^{K \eta^{u}} + \sqrt{2} I_{11}^{K \eta^{s}}\right)^{3} L_{\mu} \left[g^{\mu\nu} \right. \nonumber\\
    && + \frac{C_{K^{*}}}{g_{K^{*}}} \left(\frac{I_{11}^{K K^{*} \eta^{u}} + \sqrt{2} I_{11}^{K K^{*} \eta^{s}}}{I_{11}^{K \eta^{u}} + \sqrt{2} I_{11}^{K \eta^{s}}}\right)^{3} \frac{g^{\mu\nu} q^{2} f(q^{2}) - q^{\mu}q^{\nu}f(M_{K^{*}}^{2})}{M_{K^{*}}^{2} - q^{2} - i \sqrt{q^{2}}\Gamma_{K^{*}}} \nonumber\\
    && \left. + \frac{C_{K^{*'}}}{g_{K^{*}}} \left(\frac{I_{11}^{K K^{*'} \eta^{u}} + \sqrt{2} I_{11}^{K K^{*'} \eta^{s}}}{I_{11}^{K \eta^{u}} + \sqrt{2} I_{11}^{K \eta^{s}}}\right)^{3} \frac{g^{\mu\nu} q^{2} f(q^{2}) - q^{\mu}q^{\nu}f(M_{K^{*'}}^{2})}{M_{K^{*'}}^{2} - q^{2} - i \sqrt{q^{2}}\Gamma_{K^{*'}}} \right] F_{\nu}.
\end{eqnarray}

For the cutoff parameter equal to the one that was applied in the process $\tau \to K \pi \nu_{\tau}$ when comparing with results from Particle Data Group (PDG) \cite{Kpi} ($\Lambda_{M} = 950$ MeV), the branching fraction of this decay is in agreement with the experimental data:
\begin{eqnarray}
	Br(\tau^{-} \to K^{-} \eta \nu_{\tau}) & = & 1.56  \times 10^{-4}.
\end{eqnarray}

\section{Conclusion}
In the present and our previous work \cite{Kpi}, the processes $\tau \to K^{-} \eta \nu_{\tau}$ and $\tau \to K^{-} \pi^{0} \nu_{\tau}$ have been considered, respectively. These processes were considered earlier by using the NJL model \cite{Volkov:2017arr}. However, significant inaccuracies were admitted there: the renormalization constant $Z_{K}$ defined in (\ref{Zk}) did not take into account the splitting of the meson $K_{1}$ into two states. Besides, the transitions between the axial vector and pseudoscalar mesons were not considered, and there were a few other flaws. Therefore, the previous results should not be taken into consideration. We re-examined the process $\tau \to K^{-} \pi^{0} \nu_{\tau}$ in the work \cite{Kpi} and the process $\tau \to K^{-} \eta \nu_{\tau}$ in the present work without these inaccuracies. As a result of taking into account the meson interaction in the final state, a new cutoff parameter over the meson loop appeared. And it is the same for both the processes when comparing the results with the PDG data.

The decay considered in the present work was also studied in other theoretical papers. For instance, we would like to highlight the work \cite{Escribano:2013bca}. Different ways of taking into account final state interactions in the framework of the chiral perturbation theory with resonances were considered there. The direct use of these methods in connection with the NJL model is impossible due to the inconsistency of the respective approaches. The method of taking into account the meson interaction in the final state applied in the present work allows one to use the NJL model, although it requires going beyond its frameworks.

As mentioned in the Introduction, we have earlier calculated numerous tau lepton decays with vector mesons without taking into account the final state interaction. Nevertheless, satisfactory results were obtained. Therefore, the question of the conditions under which the final state interaction plays an important role still remains. The authors are going to consider this problem in future works.

\section*{Acknowledgements}
The authors are grateful to A. B. Arbuzov for his interest in this work and useful discussions.


\begin{thebibliography}{99}
    \bibitem{Volkov:1986zb} M.~K.~Volkov, Sov. J. Part. Nucl. 17, 186 (1986)
    \bibitem{Volkov:1993jw} M.~K.~Volkov, Phys. Part. Nucl. \textbf{24} (1993), 35-58
    \bibitem{Volkov:2005kw} M.~K.~Volkov, A.~E.~Radzhabov, Phys. Usp. 49, 551 (2006)
    \bibitem{Volkov:2017arr} M.~K.~Volkov and A.~B.~Arbuzov, Phys. Usp. 60, no. 7, 643 (2017)
    \bibitem{Volkov:2017cmv} M.~K.~Volkov, K.~Nurlan and A.~A.~Pivovarov, JETP Lett. \textbf{106} (2017) no.12, 771-774
    \bibitem{Volkov:2019yhy} M.~K.~Volkov, K.~Nurlan and A.~A.~Pivovarov, Int. J. Mod. Phys. A \textbf{34} (2019) no.24, 1950137
    \bibitem{Volkov:2019cja} M.~K.~Volkov, A.~A.~Pivovarov and K.~Nurlan, Eur. Phys. J. A \textbf{55} (2019) no.9, 165
    \bibitem{Volkov:2019udu} M.~K.~Volkov and A.~A.~Pivovarov, Pisma Zh. Eksp. Teor. Fiz. \textbf{109} (2019) no.4, 219-222 [erratum: JETP Lett. \textbf{109} (2019) no.12, 821]
    \bibitem{Volkov:2019jug} M.~K.~Volkov and A.~A.~Pivovarov, JETP Lett. \textbf{110} (2019) no.4, 237-241
    \bibitem{Volkov:2019yli} M.~K.~Volkov, A.~A.~Pivovarov and K.~Nurlan, Int. J. Mod. Phys. A \textbf{35} (2020) no.06, 2050035
    \bibitem{Volkov:2019izp} M.~K.~Volkov, A.~A.~Pivovarov and K.~Nurlan, Nucl. Phys. A \textbf{1000} (2020), 121810
    \bibitem{Volkov:2020dvz} M.~K.~Volkov, A.~B.~Arbuzov and A.~A.~Pivovarov, JETP Lett. \textbf{112} (2020) no.8, 457-462
    \bibitem{Kpi} M.~K.~Volkov and A.~A.~Pivovarov, Pisma Zh.Eksp.Teor.Fiz. \textbf{113} (2021) no.12, 777-783
    \bibitem{Volkov:1996br} M.~K.~Volkov, C.~Weiss, Phys. Rev. D 56, 221 (1997)
    \bibitem{Volkov:1996fk} M.~K.~Volkov, Phys. Atom. Nucl. 60, 1920 (1997)
    \bibitem{Bartelt:1996iv} J.~E.~Bartelt, S.~E.~Csorna, V.~Jain \textit{et al.} [CLEO], Phys. Rev. Lett. \textbf{76} (1996), 4119-4123
    \bibitem{Buskulic:1996qs} D.~Buskulic, I.~De~Bonis, D.~Decamp \textit{et al.} [ALEPH], Z. Phys. C \textbf{74} (1997), 263-273
    \bibitem{Inami:2008ar} K.~Inami, T.~Ohshima, H.~Kaji \textit{et al.} [Belle], Phys. Lett. B \textbf{672} (2009), 209-218
    \bibitem{delAmoSanchez:2010pc} P.~del Amo Sanchez, J.~P.~Lees, V.~Poireau \textit{et al.} [BaBar], Phys. Rev. D \textbf{83} (2011), 032002
    \bibitem{Li:1996md} B.~A.~Li, Phys. Rev. D \textbf{55} (1997), 1436-1452
    \bibitem{Escribano:2013bca} R.~Escribano, S.~Gonzalez-Solis and P.~Roig, JHEP \textbf{10} (2013), 039
    \bibitem{Zyla:2020zbs} P.~A.~Zyla, R.~M.~Barnett, J.~Beringer \textit{et al.} [Particle Data Group], PTEP \textbf{2020} (2020) no.8, 083C01
\end{thebibliography}
\end{document}